\documentstyle[twocolumn,aps,prl,epsf,psfig]{revtex}
\input{epsf}
\input epsf.sty
\begin{document} 

\title{Magnetic and Thermodynamic Properties of the Three-Dimensional
Periodic Anderson Hamiltonian}
\author{Carey Huscroft,$^1$ A.K. McMahan,$^2$ and R.T. Scalettar$^1$}
\address{$^1$Physics Department, University of California, Davis, CA 95616}
\address{$^2$Lawrence Livermore National Laboratory, University of California,
Livermore, California 94550}

\address{\mbox{ }} \address{\parbox{14cm}{\rm \mbox{ }\mbox{ } 
The three dimensional periodic Anderson model is studied
with Quantum Monte Carlo.
We find that the cross--over to the
Kondo singlet regime is remarkably sharp at low temperatures,
and that the behavior of magnetic correlations is  
consistently reflected in both the thermodynamics and the density of
states.  The abruptness of the transition suggests that 
energy changes associated with the 
screening of local moments by conduction electrons 
might be sufficient to drive large volume changes 
in systems where applied pressure tunes the ratio
of interband hybridization to correlation energy.
}}\address{\mbox{ }} \address{\mbox{ }}
\address{\parbox{14cm}{\rm \mbox{ }\mbox{ } PACS numbers:
71.10.Fd, 71.10.Ht, 71.27.+a}} \maketitle

\narrowtext

The problem of localized, highly correlated electrons hybridizing
with a conduction band is one of long--standing interest\cite{HFREV}.
Our understanding of the underlying physics has recently been increased
through new analytic approaches\cite{HFREV,DMFT}, and
numeric methods like Quantum Monte Carlo (QMC)\cite{QMCPAM,VEKIC}.
These techniques have emphasized 
the connection between static magnetic
properties and the dynamic response like the density of states.

However, what has been much less carefully explored by QMC
is the link to thermodynamics.
An intriguing problem for which a detailed understanding of 
the thermodynamics is essential is the ``volume--collapse''
transition in rare earth metals.
This phenomenon occurs with the application of pressure to certain
Lanthanides and gives rise to first order phase transitions 
with unusually large volume changes (14\% for Cerium and
9\% for Praseodymium)\cite{BENEDICT,JCAMD}.
Accompanying the change in volume is a change in the magnetism:
On the expanded, highly correlated, side of the transition,
the $f$ electrons have well defined moments,
while on the contracted, less correlated, side
these moments disappear or are expected to disappear.
The low-volume $\alpha$ phase of Ce is paramagnetic,
as are the early actinides which are considered to be analogs for the
collapsed rare earth phases\cite{JCAMD}.

Even the qualitative origin of this phenomenon
is still under debate.  One suggestion is that
the pressure--induced change in the ratio of the interaction
strength to bandwidth gives rise to 
a Mott transition of the 4$f$ electrons accompanied by
loss of magnetic order\cite{JOHANSSON}.
An alternate proposition is that the rapid change in
the 4$f$--valence electron coupling
leads to a ``Kondo volume collapse''\cite{ALLEN}.
In both cases, there are dramatic thermodynamic (e.g., pressure-volume)
as well as magnetic signatures of the phenomenon.

In this paper we will establish the connection between the
thermodynamics and the magnetic properties of the symmetric periodic
Anderson model (PAM) in three dimensions.
While previous efforts have focussed on the Anderson impurity 
model\cite{ALLEN,GUNNAR}, the capabilities of modern massively parallel 
computers now make feasible rigorous QMC calculations for the 
more realistic periodic model, which has so far received little
attention in three dimensions.
Our key results are:

$\bullet$  The dependence of the singlet correlation function 
on the interband hybridization
shows an increasingly sharp structure as the temperature
is lowered, indicating a very rapid cross--over between a 
regime where the $f$ sites have unscreened moments and
one in which the moments are quenched by the conduction electrons.

$\bullet$  
A sharp thermodynamic feature exists at the same
interband hybridization as this change in the singlet correlator.
To analyze this,
we introduce a new approach to the calculation of the
free energy $F$, and show it
obeys various analytic sum--rules.

$\bullet$  
The pressure difference at the transition inferred from $F$
is reasonably consistent with 
experimental pressure--volume data on Ce, Pr, and Gd, 
given the approximate 
representation of the electronic structure.

The periodic Anderson Hamiltonian is
\begin{eqnarray}
H &=& \sum_{k \sigma} \epsilon_k
d_{k\sigma}^{\dagger} d_{k\sigma}
+ \sum_{k \sigma} V_k
(d_{k\sigma}^{\dagger} f_{k\sigma}
+ f_{k\sigma}^{\dagger} d_{k\sigma}) \nonumber \\
&&+ U_{f} \sum_{i} (n_{if\uparrow} - \frac12)
(n_{if\downarrow} - \frac12) \nonumber \\
&&+ \sum_{i\sigma} \epsilon_{f} n_{if\sigma}
-\mu \sum_{i\sigma} (n_{if\sigma} +n_{id\sigma}) \, .
\label{pamham}
\end{eqnarray}   
We choose a simple cubic structure for which,
\begin{eqnarray}
\epsilon_k &=& -2 t_{dd}\,[\cos{k_x a} + \cos{k_y a} + \cos{k_z a} ] \, ,
\nonumber \\
V_k &=& -2 t_{fd}\,[\cos{k_x a} + \cos{k_y a} + \cos{k_z a} ] \, ,
\label{eq:disperse1}
\end{eqnarray}
where $a$ is the lattice constant.  The dispersion of $V_k$
reflects our choice of near--neighbor (as opposed to on--site)
hybridization of the $f$ and $d$ electrons\cite{FOOT2}.
Parameter values and temperature $T$ in this work are given in units of
$t_{dd}$.  We take $U_f=6$, consistent with the rare earths for a
reasonable choice of $t_{dd}=1$ eV, and explore a range of $t_{fd}$ and $T$ 
values.   Our choice of the ``symmetric'' PAM dictates $\mu=\epsilon_f=0$,
and thus half--filling: $\langle n_{if} \rangle = \langle n_{id} \rangle =1$.
QMC results for this model were obtained
using the determinant algorithm\cite{DET}, which
provides an exact treatment (to within statistical errors
and finite size effects) of the correlations.
There is no ``sign problem'' for the symmetric PAM,
allowing accurate simulations at low temperatures.

Figure~1 shows the temperature and $t_{fd}$ dependence of the 
singlet correlation function of near--neighbor sites {\bf i,j},
\begin{equation}
c_{fd}=\langle \vec S_{f{\bf i}} \cdot \vec S_{d{\bf j}} \rangle.
\end{equation}
Here $\vec S_{f{\bf i}} =  \left(\matrix{ f^{\dagger}_{\uparrow {\bf i}} & 
f^{\dagger}_{\downarrow {\bf i}} } \right)
\vec \sigma  \left (\matrix{ f_{\uparrow {\bf i}} \cr 
f_{\downarrow {\bf i}} }\right) $
and similarly for $\vec S_{d{\bf j}}$.
For weak interband hybridization, $c_{fd}$ is small and the
$f$ moments are unscreened by the conduction electrons.  At low
temperature, a sharp change is seen to occur at $t_{fd} \approx 0.6$
to a phase where such screening is well established.

\begin{figure}[hbt] 
\unitlength1cm \begin{picture}(8.0,7.00) \put(0.2,-0.1)
{\psfig{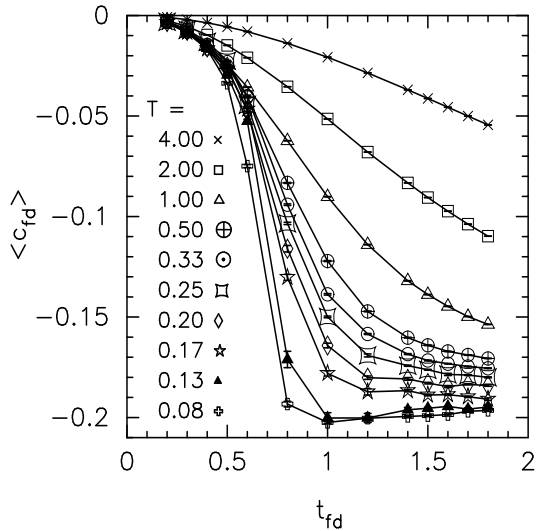}}
\end{picture} 
\vskip 0.2cm
\caption{
The singlet correlation function 
$\langle \vec S_f({\bf i}) \cdot \vec S_d({\bf j}) \rangle$
as a function of $f$--$d$ hybridization.
As the temperature is lowered, there is an
increasingly rapid switch from a small $t_{fd}$ regime 
where singlet correlations are absent to a large $t_{fd}$ 
regime where Kondo singlets are well formed.}
\end{figure}

This sharp switch is also reflected in
the energy $E$ and free energy $F$.
The difference $\Delta E(T)=E_{{\rm QMC}}(T)-E_{\rm AFHF}(T)$
of the QMC calculations relative to
antiferromagnetic Hartree-Fock (AFHF) results 
at the same temperature ($T=0.08$) is shown in Fig.~2.
To get $F=E-ST$ we fit\cite{DUFFY} the
raw data for $E_{{\rm QMC}}(T)$,
\begin{equation} 
E_{{\rm QMC}}(T) = E_0 + \sum_n c_n e^{-n\Delta/T} \, .
\end{equation} 
The number of
fitting parameters ($E_0$, $c_n$, $\Delta$) was taken to be about half
of the number of data points.  The entropy is then\cite{FOOTS0},
\begin{equation}
S(T)=S_0 + \frac{1}{T}\sum_n c_n(1+\frac{T}{n\Delta})e^{-n\Delta/T} \, .
\end{equation} 
Fig.~3 shows a plot of the resulting free energy difference 
$\Delta F(T)=F_{{\rm QMC}}(T)-F_{\rm AFHF}(T)$.
Independent fits ($E_0$, $c_n$,
$\Delta$) were performed for each $t_{fd}$, so that the smoothness of
the resultant curves in Fig.~3 is one measure of
the success of this procedure.  
Another is that our fit yields $\sum_n c_n/n\Delta$ to within $\sim
3$\% of the expected value \cite{FOOTS0} for $t_{fd}\ge 0.8$.  This sum
is smaller by $\ln{2}$ to within $\sim3$\% for $t_{fd}\leq 0.5$,
reflecting magnetic disorder of the spins below our lowest temperature
($T=0.08$) in this regime, and consequent validity of the fit only for
$T\geq 0.08$.

\vskip0.5cm
\begin{figure}[hbt] 
\unitlength1cm \begin{picture}(7.0,5.00) \put(0.0,-1.5)
{\psfig{figure=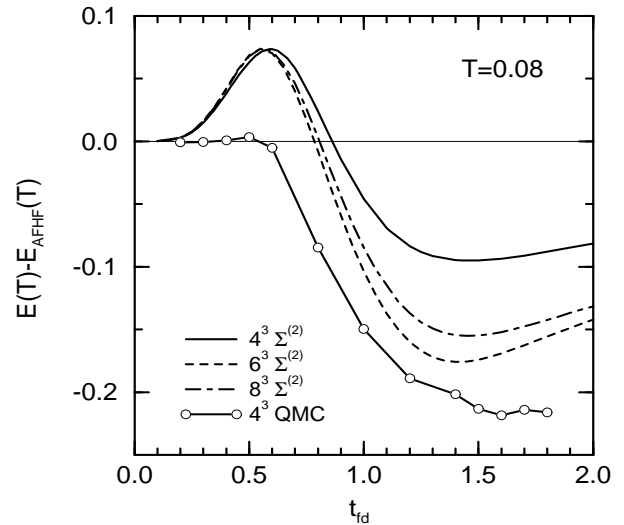,height=7.00cm,width=8.0cm,angle=0}}
\end{picture} 
\vskip2.0cm
\caption{The difference in energies between 
QMC and antiferromagnetic Hartree--Fock solutions.
At small $t_{fd}$, the AFHF energy accurately tracks
the QMC.   
However, at intermediate coupling the QMC results
break away, reflecting the failure of HF 
to pick up the singlet correlations.  
A perturbation approach (labelled $\Sigma^{(2)}$) described in the
text has some of the correct features seen in QMC.}
\end{figure}

The crucial feature in Figs.~2 and 3 is the rapid change in slope at low
temperatures of $\Delta E$ and $\Delta F$ near
$t_{fd}=0.6$.  
This behavior is hard to discern in the full
thermodynamic functions whose variation with $t_{fd}$ is $\sim 20$
times larger than seen for these difference functions.  
It arises from
the QMC results, and not from HF transitions, since the AFHF solution
is stable throughout the $t_{fd}$, $T$ region plotted here.  The size
of the present slope change is not inconsistent with the volume
collapse transitions, where one might view $\Delta F =
\min(F_1,F_2)-F_1$ with $F_1$ and $F_2$ being free energy branches
associated with the small $t_{fd}$ (large volume) and large $t_{fd}$
(small volume) phases, respectively.  Given a volume
dependence\cite{JCAMD} of $t_{fd}\sim V^{-2}$, the slope change is
related to a pressure difference by $V \Delta P/2 = -(1/2) \partial
\Delta F/\partial \ln{V} \sim \partial \Delta F/\partial \ln{t_{fd}}$.
Extrapolations of experimental pressure-volume data\cite{REexp}
into the two phase regions suggests $V \Delta P/2 \sim 0.4$, 0.5, and
1.3 eV for Ce, Pr, and Gd, respectively.  The low-$T$ slope change in
Figs.~2 and 3 is
$\partial \Delta F/\partial \ln{t_{fd}} = 0.2$--0.3 eV,
which given the crudeness of the present
representation of the rare earth valence electrons is
reasonably consistent.
                                
The QMC calculations were carried out for a $4^3$-site
lattice.  As a systematic exploration of system size for these
three--dimensional calculations would be prohibitive,
we have used a
second--order self-energy approach to estimate the
size effects, as well as to explore what analytic approximations might be more
suitable than HF to capture the thermodynamics.
The solid ($4^3$), dash--dot ($6^3$), and dash ($8^3$) 
curves in Fig.~2 labelled
$\Sigma^{(2)}$ were obtained from a finite--T version of the
self-energy approach of Steiner {\it et.~al.}\cite{steiner91}. 
The Dyson equation for the interacting
Green's function matrix $G_{\bf k}$ is solved based on a second order (in $U_f$)
expression for the self--energy $\Sigma^{(2)}_{\bf k}(\{G^{(0)}_{\bf
k}\})$, determined from the paramagnetic HF result $G^{(0)}_{\bf k}$.
The trend in the $\Delta E_\Sigma^{(2)}$ curves for 
periodic clusters of $4^3$, $6^3$, and $8^3$ sites suggests that
finite size effects\cite{FOOT3} do not alter the qualitative physics and,
indeed, move the position of the
transition in the correct direction for comparison with experiment,
namely to higher values of $\partial \Delta F/\partial \ln{t_{fd}}$.

\vskip0.5cm
\begin{figure}[hbt] 
\unitlength1cm \begin{picture}(7.0,5.00) \put(0.0,-1.5)
{\psfig{figure=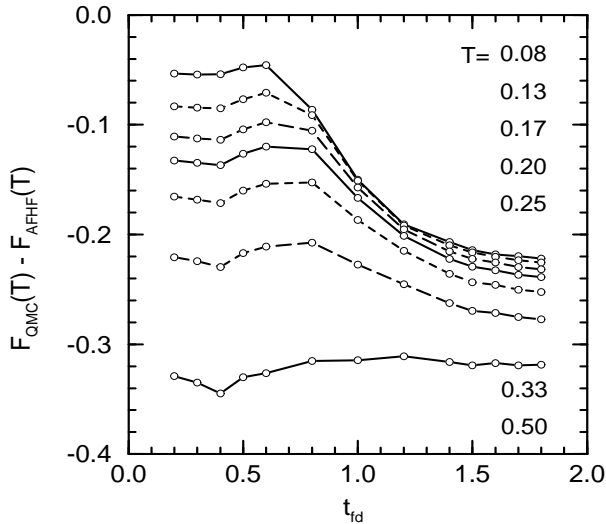,height=7.00cm,width=8.0cm,angle=0}}
\end{picture} 
\vskip2.0cm
\caption{The difference in free energies between 
QMC and antiferromagnetic Hartree--Fock solutions.
At strong coupling (small $t_{fd}$) the 
agreement in the free energy is good
apart from an overall shift of $T\ln{2}$ associated
with the tendency of HF to overestimate the magnetic order. As in
Fig.~2, at intermediate coupling $\Delta F$ becomes sizeable.}
\end{figure}

There is striking consistency between the singlet correlations in
Fig.~1 and the energy and free energy differences in Figs.~2,3.  
In all cases there
is a rather abrupt switch in low--T behavior across $t_{fd} \sim 0.6$,
which anneals with increasing temperature.   The
anomalies are largely gone above $T\sim 0.5$, an upper bound for what
might be a critical temperature in the present model.  The actual
critical temperature will reflect competition between effects like
these in $\Delta F$ and the volume dependence of a realistic
generalization of $F_{\rm AFHF}$.  An important term in $\Delta F$ is
the QMC entropy, which reflects disordered spins for small $t_{fd}$ at
the lowest temperature $T=0.08$, in contrast to both larger $t_{fd}$
values at this temperature, as well as the stable AFHF solution
throughout the range plotted in Fig.~3, where the entropy is
approximately minimal.  Consequently, $\Delta F$ includes a $-T
\ln{2}$ entropy term at small $t_{fd}$, but not at large $t_{fd}$,
which serves to level out the $\Delta F$ curves as
temperature is increased.

Besides singlet formation, magnetic ordering of the local moments is
a generic feature of the PAM.  Indeed, our
calculations of the $f$--$f$ structure factor suggest a strong tendency
for the $f$ moments to order antiferromagnetically at low temperatures
with a maximal ordering temperature 
in the vicinity of $t_{fd}=0.8$.  
Further insight into
the relation between AF, singlet formation, and the thermodynamics can
be obtained by computing the heat capacity
$C(T)=dE(T)/dT$\cite{JCAMD}. 
We find a low--temperature peak 
similar to recent work on the two-dimensional Hubbard model\cite{DUFFY},
with an area $\int dT C(T)/T$ of $\ln{2}$ at small $t_{fd}$,
which, however, washes out with decreasing area at large $t_{fd}$.  The peak has
only minor impact on the slope change discussed above for $\Delta
E_{\rm QMC}$ and $\Delta F_{\rm QMC}$ as functions of $t_{fd}$
\cite{FOOT5}.

A more complete picture of the PAM
is given by the density of states,
$N_f(\omega)$, which we obtain using
the Maximum Entropy method\cite{MAXENT} 
to perform the analytic continuation of the imaginary
time Greens function computed in QMC.
The results for different $t_{fd}$ at fixed
$T=0.2$ are shown in Fig.~4.
$N_{f}(\omega)$ evolves from a structure with
upper and lower Hubbard bands separated by a gap 
$U_{f}$ at small $t_{fd}$ to a regime where broadened remnants of
these bands are still evident but additional resonant peaks characteristic
of Kondo singlet formation have also developed.  
As $t_{fd}$ is increased, central resonances
appear and are sharpest at $t_{fd}\approx 0.6$, indicating
the onset of singlet formation.  Further increase in $t_{fd}$ enhances
the weight in this central region at the expense of the Hubbard
sidebands.  

The precise nature of the gap in the density of states 
at the Fermi surface, $\omega=0$,
is still open to interpretation.
For the half--filled, single band Hubbard Hamiltonian, $N(\omega)$
has a similar gap which
evolves continuously from predominantly Mott--Hubbard character,
for $U >> W$, to a Slater gap associated with antiferromagnetic order,
for $U<<W$.  Similarly, the two--band model
considered here has a Mott gap at
small $t_{fd}$, while the gap at larger $t_{fd}$ could originate
either as a result of long range antiferromagnetic
order on the $f$ sites, or, alternately,
reflect a ``coherence gap'' 
associated with singlet formation.
The competition between these two latter effects
on $N(\omega)$ is well documented in a lower dimension\cite{VEKIC}.
Here, studies of the $f$--$f$ correlation function show no signs of
AF long-range order at $t_{fd} = 0.6$ and $T=0.2 > T_{{\rm Neel}}$, 
which suggests 
these resonances signal singlet formation, not AFLRO.
Analytic continuation of two particle Green's functions, 
like the magnetic susceptibility, will lend further insight into
this question, but is very difficult and remains to be done.

\begin{figure}[hbt] \unitlength1cm 
\begin{picture}(7.0,7.00) \put(0.0,-0.1)
{\psfig{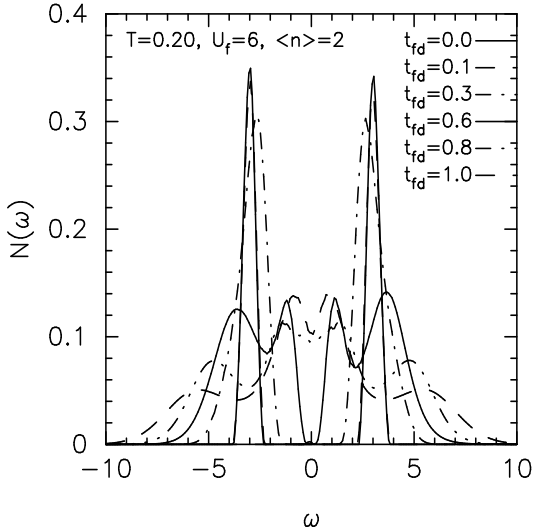}}
\end{picture} 
\vskip 0.2cm
\caption{The $f$--band density of states for different $f$--$d$
hybridization.  For weak hybridization,  there are 
peaks at $\pm U_{f}/2$.
These broaden with increasing $t_{fd}$, and a Kondo
resonance develops.  The curves for the two smallest
$t_{fd}$ values have been reduced by a factor of two for display purposes.}
\end{figure}

In this paper we have shown that there is a striking consistency between
the location of sharp cross--overs in the singlet magnetic
and thermodynamic properties of the three--dimensional
periodic Anderson model.  The $f$ density
of states shows a structure expected to arise
from singlet correlations.  Finally, estimates of the associated change
in free energy are of the same order of magnitude as observed in the
rare earth volume collapse transitions.

Two important issues remain open.  The first is 
the extension to Hamiltonians with the full
rare earth orbital complexity.
Initial studies of how the Mott transition varies with band
degeneracy in the Hubbard model, and other issues, 
already exist\cite{KOTLIAR}
within approximate numerical approaches
like dynamical mean field theory\cite{DMFT}.
The second, related, issue concerns band filling.  
Studies with many $f$ orbitals will require working away
from the symmetric point. 

Work at UCD was supported in part by an Accelerated Strategic Computing
Initiative grant and by the LLNL Materials Research Institute; 
that at LLNL, by
the U.S. Department of Energy under Contract No.~W--7405--Eng--48.  The
QMC calculations were performed on the ASCI Blue-Pacific and Red
platforms.


\begin{references}
\bibitem{FOOT2}
With a momentum independent $V_k$, the PAM is a band insulator 
at half--filling at $U_f=0$.
Our choice has no such hybridization gap and is therefore more
representative of the rare earth metals.
\bibitem{FOOTS0}
The constant $S_0 \equiv 4\ln{2}-\sum_n c_n/n\Delta$, and should equal 
$(2N_0/N)\ln{2}$ where $N_0$ is the number of discrete
$k$ points which have $\epsilon_k=V_k=0$ on a finite lattice.
$N_0/N \rightarrow 0$ as $N \rightarrow \infty.$
\bibitem{FOOT5}
One way to see this is from Fig.~2.
The paramagnetic $\Sigma^{(2)}$ calculation in Fig.~2 has no low-$T$
peak in $C(T)$ yet yields structure similar to the full QMC results.
Moreover, with guidance from the temperature dependence of
$E_{\Sigma^{(2)}}(T)$, one may extrapolate low-$T$ ``QMC'' energies for
which the $C(T)$ peak has been artificially removed.  These modified
QMC energies yield a $\Delta E_{\rm QMC}$ curve only slightly different
from that in Fig.~2.
\end{references}
\end{document}